\documentclass{article}
\usepackage{amssymb}
\usepackage{amsmath}
 \allowdisplaybreaks
\begin{document}

\title{On the uniqueness of $D=11$ interactions among a graviton, a massless gravitino and a three-form.\\
II: Three-form and gravitini}

\author{E. M. Cioroianu\thanks{ e-mail address:
manache@central.ucv.ro}, E. Diaconu\thanks{ e-mail address:
ediaconu@central.ucv.ro}, S. C. Sararu\thanks{ e-mail
address: scsararu@central.ucv.ro}\\
Faculty of Physics, University of Craiova,\\ 13 Al. I. Cuza Street
Craiova, 200585, Romania}

\maketitle

\begin{abstract}
The interactions that can be introduced between a massless
Rarita-Schwinger field and an Abelian three-form gauge field in
eleven spacetime dimensions are analyzed in the context of the
deformation of the \textquotedblleft free\textquotedblright\
solution of the master equation combined with local BRST cohomology.
Under the hypotheses of smoothness of the interactions in the
coupling constant, locality, Poincar\'{e} invariance, Lorentz
covariance, and the presence of at most two derivatives in the
Lagrangian of the interacting theory (the same number of derivatives
like in the free Lagrangian), we prove that there are neither
cross-couplings nor self-interactions for the gravitino in $D=11$.
The only possible term that can be added to the deformed solution to
the master equation is nothing but a generalized Chern-Simons term
for the three-form gauge field, which brings contributions to the
deformed Lagrangian, but does not modify the original, Abelian gauge
transformations.


\end{abstract}

\section{Introduction}

It is known that the field content of $D=11$, $N=1$ supergravity is
remarkably simple; it consists of a graviton, a massless Majorana
spin-$3/2$ field, and a three-form gauge field. The analysis of all
possible interactions in $D=11$ related to this field content
necessitates the study of cross-couplings involving each pair of
these sorts of fields and then the construction of simultaneous
interactions among all the three fields. With this purpose in mind,
in Ref.~\cite{SUGRAI} we have obtained all consistent interactions
that can be added to a free theory describing a massless spin-two
field and an Abelian three-form gauge field in eleven spacetime
dimensions. Here, we develop the second step of our approach and
analyze the consistent eleven-dimensional interactions that can be
introduced between a massless Rarita-Schwinger field and an Abelian
three-form gauge field. Our main result is that under the hypotheses
of smoothness of the interactions in the coupling constant,
locality, Poincar\'{e} invariance, Lorentz covariance, and the
presence of at most two derivatives in the Lagrangian of the
interacting theory (the same number of derivatives like in the free
Lagrangian) there are neither cross-couplings nor self-interactions
for the gravitino in $D=11$. The only possible term that can be
added to the deformed solution to the master equation is nothing but
a generalized Chern-Simons term for the three-form gauge field,
which brings contributions to the deformed Lagrangian, but does not
modify the original, Abelian gauge transformations. Our result does
not contradict the presence in the Lagrangian of $D=11$, $N=1$ SUGRA
of a quartic vertex expressing self-interactions among the
gravitini. We will see in Refs.~\cite{SUGRAIII} and~ \cite{SUGRAIV}
that this vertex, which appears at order two in the coupling
constant, is due to the \textit{simultaneous} presence of gravitini,
three-form, \textit{and} graviton.

\section{Free model: Lagrangian formulation and BRST symmetry}

Our starting point is represented by a free model, whose Lagrangian action
is written like the sum between the standard action of an Abelian three-form
gauge field and that of a massless Rarita-Schwinger field in eleven
spacetime dimensions
\begin{eqnarray}
S_{0}^{\mathrm{L}}\left[ A_{\mu \nu \rho },\psi _{\mu }\right] &=&\int
d^{11}x\left( -\frac{1}{2\cdot 4!}F_{\mu \nu \rho \lambda }F^{\mu \nu \rho
\lambda }-\frac{\mathrm{i}}{2}\bar{\psi}_{\mu }\gamma ^{\mu \nu \rho
}\partial _{\nu }\psi _{\rho }\right)  \notag \\
&\equiv &\int d^{11}x\left( \mathcal{L}_{0}^{\mathrm{A}}+\mathcal{L}_{0}^{%
\mathrm{\psi }}\right) ,  \label{3frs1}
\end{eqnarray}%
where $F_{\mu \nu \rho \lambda }$ denotes the field strength of the
thee-form gauge field ($F_{\mu \nu \rho \lambda }=\partial _{\lbrack
\mu }A_{\nu \rho \lambda ]}$). We maintain the antisymmetrization
convention explained in part I \cite{SUGRAI} and work with that
representation of the Clifford algebra
\begin{equation}
\gamma _{\mu }\gamma _{\nu }+\gamma _{\nu }\gamma _{\mu }=2\sigma _{\mu \nu }%
\mathbf{1}  \label{3frs2}
\end{equation}%
for which all the $\gamma $ matrices are purely imaginary. In addition, we
take $\gamma _{0}$ to be Hermitian and antisymmetric and $\left( \gamma
_{i}\right) _{i=\overline{1,10}}$ anti-Hermitian and symmetric
\begin{eqnarray}
\left( \gamma _{\mu }\right) ^{\ast } &=&-\gamma _{\mu },  \label{3frs3} \\
\gamma _{\mu }^{\intercal } &=&-\gamma _{0}\gamma _{\mu }\gamma _{0},\qquad
\mu =\overline{0,10}.  \label{3frs4}
\end{eqnarray}%
The operations of Dirac and respectively Majorana conjugation are defined as
usually via the relations
\begin{eqnarray}
\bar{\psi}_{\mu } &=&\left( \psi _{\mu }\right) ^{\dagger }\gamma _{0},
\label{3frs5} \\
\psi ^{c} &=&\left( \mathcal{C}\psi \right) ^{\intercal },  \label{3frs6}
\end{eqnarray}%
where the charge conjugation matrix is
\begin{equation}
\mathcal{C}=-\gamma _{0}.  \label{3frs7}
\end{equation}%
In what follows we use the notations
\begin{equation}
\gamma _{\mu _{1}\cdots \mu _{k}}=\frac{1}{k!}\sum\limits_{\Theta \in \Sigma
_{k}}\left( -\right) ^{\Theta }\gamma _{\mu _{\Theta (1)}}\gamma _{\mu
_{\Theta (2)}}\cdots \gamma _{\mu _{\Theta (k)}},  \label{fie3}
\end{equation}%
where $\Sigma _{k}$ represents the set of permutations of the numbers $%
\left\{ 1,2,\ldots ,k\right\} $ and $\left( -\right) ^{\Theta }$ is the
signature of a given permutation $\Theta $. We will need the Fierz
identities specific to $D=11$%
\begin{equation}
\gamma _{\mu _{1}\cdots \mu _{p}}\gamma ^{\nu _{1}\cdots \nu
_{q}}=\sum\limits_{p+q-11\leq 2k\leq 2M}\delta _{\lbrack \mu _{p}}^{[\nu
_{1}}\delta _{\mu _{p-1}}^{\nu _{2}}\cdots \delta _{\mu _{p-k+1}}^{\nu
_{k}}\gamma _{\mu _{1}\cdots \mu _{p-k}]}^{\qquad \qquad \nu _{k+1}\cdots
\nu _{q}]},  \label{fie1}
\end{equation}%
where $M=\min (p,q)$ and also the development of a complex, spinor-like
matrix $N$ in terms of the basis $\left\{ \mathbf{1},\gamma _{\mu },\gamma
_{\mu \nu },\gamma _{\mu \nu \rho },\gamma _{\mu \nu \rho \lambda },\gamma
_{\mu \nu \rho \lambda \sigma }\right\} $%
\begin{equation}
N=\frac{1}{32}\sum\limits_{k=0}^{5}\left( -\right) ^{k(k-1)/2}\frac{1}{k!}%
\mathrm{Tr}\left( \gamma ^{\mu _{1}\cdots \mu _{k}}N\right) \gamma _{\mu
_{1}\cdots \mu _{k}}.  \label{fie2}
\end{equation}%
The theory described by action (\ref{3frs1}) possesses an Abelian,
off-shell, second-order reducible generating set of gauge transformations
\begin{equation}
\delta _{\varepsilon }A_{\mu \nu \rho }=\partial _{\lbrack \mu }\varepsilon
_{\nu \rho ]},\qquad \delta _{\varepsilon }\psi _{\mu }=\partial _{\mu
}\varepsilon .  \label{3frs8}
\end{equation}%
Related to the gauge parameters, $\varepsilon _{\mu \nu }$\ are
bosonic and completely antisymmetric and $\varepsilon $ is a
fermionic Majorana spinor. The fact that the gauge transformations
of the three-form gauge field are off-shell, second-order reducible
is treated in more detail in Ref.~\cite{SUGRAI}.

In order to construct the BRST symmetry for (\ref{3frs1}) we introduce the
field, ghost, and antifield spectra
\begin{eqnarray}
\Phi ^{\Delta _{0}} &=&\left( A_{\mu \nu \rho },\psi _{\mu }\right) ,\qquad
\Phi _{\Delta _{0}}^{\ast }=\left( A^{\ast \mu \nu \rho },\psi ^{\ast \mu
}\right) ,  \label{3frs11a} \\
\eta ^{\Delta _{1}} &=&\left( C_{\mu \nu },\xi \right) ,\qquad \eta _{\Delta
_{1}}^{\ast }=\left( C^{\ast \mu \nu },\xi ^{\ast }\right) ,  \label{3frs11b}
\\
\eta ^{\Gamma _{2}} &=&\left( C_{\mu }\right) ,\qquad \eta _{\Gamma
_{2}}^{\ast }=\left( C^{\ast \mu }\right) ,  \label{3frs11c} \\
\eta ^{\Gamma _{3}} &=&\left( C\right) ,\qquad \eta _{\Gamma _{3}}^{\ast
}=\left( C^{\ast }\right) .  \label{3frs11d}
\end{eqnarray}%
The fermionic ghosts $C_{\mu \nu }$ correspond to the gauge parameters of
the three-form, $\varepsilon _{\mu \nu }$, the bosonic ghost $\xi $ is
associated with the gauge parameter $\varepsilon $, while the bosonic ghosts
for ghosts $\eta ^{\Gamma _{2}}$ and the fermionic ghost for ghost for ghost
$\eta ^{\Gamma _{3}}$ are due to the first- and respectively second-order
reducibility of the gauge transformations from the three-form sector. The
star variables represent the antifields of the corresponding fields/ghosts.
The antifields of the Rarita-Schwinger field are bosonic, purely imaginary
spinors. Since the gauge generators of the free theory under study are field
independent, it follows that the BRST differential decomposes again like in Ref.~%
\cite{SUGRAI}%
\begin{equation}
s=\delta +\gamma ,  \label{3frs12}
\end{equation}%
where $\delta $ represents the Koszul-Tate differential and $\gamma
$ stands for the exterior derivative along the gauge orbits. (More
details of the various graduations of the BRST generators can be
found in Ref.~\cite{SUGRAI}.) In agreement with the standard rules
of the BRST formalism, the degrees of the BRST generators are valued
like
\begin{eqnarray}
\mathrm{agh}\left( \Phi ^{\Delta _{0}}\right) &=&\mathrm{agh}\left( \eta
^{\Delta _{1}}\right) =\mathrm{agh}\left( \eta ^{\Gamma _{2}}\right) =%
\mathrm{agh}\left( \eta ^{\Gamma _{3}}\right) =0,  \label{3frs13a} \\
\mathrm{agh}\left( \Phi _{\Delta _{0}}^{\ast }\right) &=&1,\quad \mathrm{agh}%
\left( \eta _{\Delta _{1}}^{\ast }\right) =2,\quad \mathrm{agh}\left( \eta
_{\Gamma _{2}}^{\ast }\right) =3,\quad \mathrm{agh}\left( \eta _{\Gamma
_{3}}^{\ast }\right) =4,  \label{3frs13b} \\
\mathrm{pgh}\left( \Phi ^{\Delta _{0}}\right) &=&0,\quad \mathrm{pgh}\left(
\eta ^{\Delta _{1}}\right) =1,\quad \mathrm{pgh}\left( \eta ^{\Gamma
_{2}}\right) =2,\quad \mathrm{pgh}\left( \eta ^{\Gamma _{3}}\right) =3,
\label{3frs13c} \\
\mathrm{pgh}\left( \Phi _{\Delta _{0}}^{\ast }\right) &=&\mathrm{pgh}\left(
\eta _{\Delta _{1}}^{\ast }\right) =\mathrm{pgh}\left( \eta _{\Gamma
_{2}}^{\ast }\right) =\mathrm{pgh}\left( \eta _{\Gamma _{3}}^{\ast }\right)
=0.  \label{3frs13d}
\end{eqnarray}%
The actions of the differentials $\delta $ and $\gamma $ on the generators
from the BRST complex are given by
\begin{eqnarray}
\delta A^{\ast \mu \nu \rho } &=&\tfrac{1}{3!}\partial _{\lambda }F^{\mu \nu
\rho \lambda },\qquad \delta \psi ^{\ast \mu }=-\mathrm{i}\partial _{\rho }%
\bar{\psi}_{\lambda }\gamma ^{\rho \lambda \mu },  \label{3frs14a} \\
\delta C^{\ast \mu \nu } &=&-3\partial _{\rho }A^{\ast \mu \nu \rho },\qquad
\delta \xi ^{\ast }=\partial _{\mu }\psi ^{\ast \mu },  \label{3frs14b} \\
\delta C^{\ast \mu } &=&-2\partial _{\nu }C^{\ast \mu \nu },\qquad \delta
C^{\ast }=-\partial _{\mu }C^{\ast \mu },  \label{3frs14c} \\
\delta \left( \Phi ^{\Delta _{0}}\right) &=&\delta \left( \eta ^{\Delta
_{1}}\right) =\delta \left( \eta ^{\Gamma _{2}}\right) =\delta \left( \eta
^{\Gamma _{3}}\right) =0,  \label{3frs14d} \\
\gamma \left( \Phi _{\Delta _{0}}^{\ast }\right) &=&\gamma \left( \eta
_{\Delta _{1}}^{\ast }\right) =\gamma \left( \eta _{\Gamma _{2}}^{\ast
}\right) =\gamma \left( \eta _{\Gamma _{3}}^{\ast }\right) =0,
\label{3frs14e} \\
\gamma A_{\mu \nu \rho } &=&\partial _{\lbrack \mu }C_{\nu \rho ]},\qquad
\gamma \psi _{\mu }=\partial _{\mu }\xi ,  \label{3frs14f} \\
\gamma C_{\mu \nu } &=&\partial _{\lbrack \mu }C_{\nu ]},\qquad \gamma \xi
=0,  \label{3frs14g} \\
\gamma C_{\mu } &=&\partial _{\mu }C,\qquad \gamma C=0.  \label{3frs14h}
\end{eqnarray}%
In this case the anticanonical action of the BRST symmetry, $s\cdot =\left(
\cdot ,S^{\mathrm{A,\psi }}\right) $, is realized via a solution to the
master equation $\left( S^{\mathrm{A,\psi }},S^{\mathrm{A,\psi }}\right) =0$
that reads as
\begin{eqnarray}
&&S^{\mathrm{A,\psi }}=S_{0}^{\mathrm{L}}\left[ A_{\mu \nu \rho },\psi _{\mu
}\right] +\int d^{11}x\left( \psi ^{\ast \mu }\partial _{\mu }\xi +A^{\ast
\mu \nu \rho }\partial _{\lbrack \mu }C_{\nu \rho ]}\right.  \notag \\
&&\left. +C^{\ast \mu \nu }\partial _{\lbrack \mu }C_{\nu ]}+C^{\ast \mu
}\partial _{\mu }C\right) .  \label{3frs15}
\end{eqnarray}

\section{Consistent interactions between an Abelian three-form gauge field
and a Rarita-Schwinger spinor}

The aim of this section is to investigate the cross-couplings that can be
introduced between an Abelian three-form gauge field and a massless
Rarita-Schwinger field in $D=11$. This matter is addressed, like in Ref.~\cite%
{SUGRAI}, in the context of the antifield-BRST deformation procedure. Very
briefly, this means that we will associate with (\ref{3frs15}) a deformed
solution
\begin{eqnarray}
S^{\mathrm{A,\psi }} &\rightarrow &\bar{S}^{\mathrm{A,\psi }}=S^{\mathrm{%
A,\psi }}+\lambda S_{1}^{\mathrm{A,\psi }}+\lambda ^{2}S_{2}^{\mathrm{A,\psi
}}+\cdots  \notag \\
&=&S^{\mathrm{A,\psi }}+\lambda \int d^{D}x\,a^{\mathrm{A,\psi }}+\lambda
^{2}\int d^{D}x\,b^{\mathrm{A,\psi }}+\cdots ,  \label{3frs16}
\end{eqnarray}%
which is the BRST generator of the interacting theory, $\left( \bar{S}^{%
\mathrm{A,\psi }},\bar{S}^{\mathrm{A,\psi }}\right) =0$, such that the
components of $\bar{S}^{\mathrm{A,\psi }}$ are restricted to satisfy the
tower of equations:
\begin{eqnarray}
\left( S^{\mathrm{A,\psi }},S^{\mathrm{A,\psi }}\right) &=&0,
\label{3frs18a} \\
2\left( S_{1}^{\mathrm{A,\psi }},S^{\mathrm{A,\psi }}\right) &=&0,
\label{3frs18b} \\
2\left( S_{2}^{\mathrm{A,\psi }},S^{\mathrm{A,\psi }}\right) +\left( S_{1}^{%
\mathrm{A,\psi }},S_{1}^{\mathrm{A,\psi }}\right) &=&0,  \label{3frs18c} \\
\left( S_{3}^{\mathrm{A,\psi }},S^{\mathrm{A,\psi }}\right) +\left( S_{1}^{%
\mathrm{A,\psi }},S_{2}^{\mathrm{A,\psi }}\right) &=&0,  \label{3frs18d} \\
&&\vdots  \notag
\end{eqnarray}%
The interactions are obtained under the same assumptions like in Ref.~\cite%
{SUGRAI}: smoothness, locality, Lorentz covariance, Poincar\'{e} invariance,
and preservation of the number of derivatives on each field (derivative
order assumption). The `derivative order assumption' means here that the
following two requirements are simultaneously satisfied: (i) the derivative
order of the equations of motion on each field is the same for the free and
respectively for the interacting theory; (ii) the maximum number of
derivatives in the interaction vertices is equal to two, i.e. the maximum
number of derivatives from the free Lagrangian.

\subsection{First-order deformation}

Initially, we construct the first-order deformation of the solution to the
master equation, $S_{1}^{\mathrm{A,\psi }}$, as solution to equation (\ref%
{3frs18b}). If we make the notation $S_{1}^{\mathrm{A,\psi }}=\int
d^{11}x\,a^{\mathrm{A,\psi }}$, with $a^{\mathrm{A,\psi }}$ a local function
($\mathrm{gh}\left( a\right) =0$, $\varepsilon \left( a\right) =0$), then (%
\ref{3frs18b}) takes the local form
\begin{equation}
sa^{\mathrm{A,\psi }}=\partial _{\mu }m^{\mu },  \label{3frs19}
\end{equation}%
which shows that the nonintegrated density of the first-order deformation
pertains to the local cohomology of the BRST differential in ghost number
zero, $a^{\mathrm{A,\psi }}\in H^{0}\left( s|d\right) $. In order to analyze
equation (\ref{3frs18b}) we act like in Ref.~\cite{SUGRAI}: we develop $a^{%
\mathrm{A,\psi }}$ according to the antighost number
\begin{equation}
a^{\mathrm{A,\psi }}=\sum\limits_{i=0}^{I}a_{i}^{\mathrm{A,\psi }},\;\mathrm{%
agh}\left( a_{i}^{\mathrm{A,\psi }}\right) =i  \label{3frs21}
\end{equation}%
and obtain in the end that equation (\ref{3frs19}) becomes equivalent to the
tower of equations
\begin{eqnarray}
\gamma a_{I}^{\mathrm{A,\psi }} &=&\partial _{\mu }\overset{\left( I\right) }%
{m}^{\mu },  \label{3frs22a} \\
\delta a_{I}^{\mathrm{A,\psi }}+\gamma a_{I-1}^{\mathrm{A,\psi }}
&=&\partial _{\mu }\overset{\left( I-1\right) }{m}^{\mu },  \label{3frs22b}
\\
\delta a_{i}^{\mathrm{A,\psi }}+\gamma a_{i-1}^{\mathrm{A,\psi }}
&=&\partial _{\mu }\overset{\left( i-1\right) }{m}^{\mu },\qquad 1\leq i\leq
I-1,  \label{3frs22c}
\end{eqnarray}%
where, moreover, equation (\ref{3frs22a}) can be replaced in strictly
positive antighost numbers by
\begin{equation}
\gamma a_{I}^{\mathrm{A,\psi }}=0,\quad I>0.  \label{3frs23}
\end{equation}%
The nontriviality of the first-order deformation $a^{\mathrm{A,\psi }}$ is
thus translated at its highest antighost number component into the
requirement that $a_{I}^{\mathrm{A,\psi }}\in H^{I}\left( \gamma \right) $,
where $H^{I}\left( \gamma \right) $ denotes the cohomology of the exterior
longitudinal derivative $\gamma $ in pure ghost number equal to $I$. So, in
order to solve equation (\ref{3frs19}) we need to compute the cohomology of $%
\gamma $, $H\left( \gamma \right) $, and, as it will be made clear below,
also the local cohomology of $\delta $ in pure ghost number zero, $H\left(
\delta |d\right) $.

Using the results on the cohomology of the exterior longitudinal
differential for an Abelian three-form gauge field computed in
Ref.~\cite{SUGRAI} as well as definitions
(\ref{3frs14e})--(\ref{3frs14h}), we can state that
the most general solution to (\ref{3frs23}) can be written, up to $\gamma $%
-exact contributions, as
\begin{equation}
a_{I}^{\mathrm{A,\psi }}=\tilde{\alpha}_{I}\left( \left[ F_{\mu \nu \rho
\lambda }\right] ,\left[ \partial _{\lbrack \mu }\psi _{\nu ]}\right] ,\left[
\chi _{\Delta }^{\ast }\right] \right) \omega ^{I}\left( C,\xi \right) ,
\label{3frs25}
\end{equation}%
where $\chi _{\Delta }^{\ast }=\left\{ \Phi _{\Delta _{0}}^{\ast },\eta
_{\Delta _{1}}^{\ast },\eta _{\Gamma _{2}}^{\ast },\eta _{\Gamma _{3}}^{\ast
}\right\} $ and $\omega ^{I}$ denotes the elements with pure ghost number $I$
of a basis in the space of polynomials in the corresponding ghosts. The
objects $\tilde{\alpha}_{I}$ (with $\mathrm{agh}\left( \tilde{\alpha}%
_{I}\right) =I$) are nontrivial elements of $H^{0}\left( \gamma \right) $,
known as \textquotedblleft invariant polynomials\textquotedblright . They
are in fact polynomials in the antifields $\chi _{\Delta }^{\ast }$, in the
field strength of the three-form $F_{\mu \nu \rho \lambda }$, in the
antisymmetrized first-order derivatives of the Rarita-Schwinger fields $%
\partial _{\lbrack \mu }\psi _{\nu ]}$ as well as in their subsequent
derivatives. Just like in Ref.~\cite{SUGRAI}, it can be shown that a
necessary condition for the existence of (nontrivial) solutions
$a_{I-1}$ is that the invariant polynomials $\tilde{\alpha}_{I}$ are
(nontrivial) objects from the local cohomology of the Koszul-Tate
differential $H\left( \delta |d\right) $ in antighost number $I>0$
and in pure ghost number zero. Using the fact that the Cauchy order
of the free theory under study is equal to four together with the
general result according to which the local cohomology of the
Koszul-Tate differential in pure ghost number zero is trivial in
antighost numbers strictly greater than its Cauchy order, we can
state that
\begin{equation}
H_{J}\left( \delta |d\right) =0\qquad \mathrm{for\;all\;}J>4.  \label{3frs27}
\end{equation}%
On the other hand, it can be shown that any invariant polynomial $\tilde{%
\alpha}_{J}$ that is trivial in $H_{J}\left( \delta |d\right) $ with $J\geq
4 $ can be taken to be trivial also in the invariant characteristic
cohomology in antighost number $J$, $H_{J}^{\mathrm{inv}}\left( \delta
|d\right) $:
\begin{equation}
\left( \tilde{\alpha}_{J}=\delta \tilde{b}_{J+1}+\partial _{\mu }\tilde{c}%
_{J}^{\mu },\;\mathrm{agh}\left( \tilde{\alpha}_{J}\right) =J\geq 4\right)
\Rightarrow \tilde{\alpha}_{J}=\delta \tilde{\beta}_{J+1}+\partial _{\mu }%
\tilde{\gamma}_{J}^{\mu },  \label{3frs28}
\end{equation}%
with both $\tilde{\beta}_{J+1}$ and $\tilde{\gamma}_{J}^{\mu }$ invariant
polynomials. Results (\ref{3frs27}) and (\ref{3frs28}) yield the conclusion
that
\begin{equation}
H_{J}^{\mathrm{inv}}\left( \delta |d\right) =0\qquad \mathrm{for\;all\;}J>4.
\label{3frs29}
\end{equation}%
It can be shown that the spaces $\left( H_{J}\left( \delta |d\right) \right)
_{J\geq 2}$ and $\left( H_{J}^{\mathrm{inv}}\left( \delta |d\right) \right)
_{J\geq 2}$ are spanned by
\begin{eqnarray}
H_{4}\left( \delta |d\right) ,H_{4}^{\mathrm{inv}}\left( \delta |d\right)
&:&\left( C^{\ast }\right) ,  \label{3frs30a} \\
H_{3}\left( \delta |d\right) ,H_{3}^{\mathrm{inv}}\left( \delta |d\right)
&:&\left( C^{\ast \mu }\right) ,  \label{3frs30b} \\
H_{2}\left( \delta |d\right) ,H_{2}^{\mathrm{inv}}\left( \delta |d\right)
&:&\left( C^{\ast \mu \nu },\xi ^{\ast }\right) .  \label{3frs30c}
\end{eqnarray}%
These results on $H\left( \delta |d\right) $ and $H^{\mathrm{inv}}\left(
\delta |d\right) $ in strictly positive antighost numbers are important
because they allow the elimination of all pieces with $I>4$ from (\ref%
{3frs21}).

In the case $I=4$ the nonintegrated density of the first-order deformation $%
a^{\mathrm{A,\psi }}$, (\ref{3frs21}), becomes
\begin{equation}
a^{\mathrm{A,\psi }}=a_{0}^{\mathrm{A,\psi }}+a_{1}^{\mathrm{A,\psi }%
}+a_{2}^{\mathrm{A,\psi }}+a_{3}^{\mathrm{A,\psi }}+a_{4}^{\mathrm{A,\psi }}.
\label{3frs31}
\end{equation}%
We can further decompose $a^{\mathrm{A,\psi }}$ in a natural manner, as a
sum between three kinds of deformations
\begin{equation}
a^{\mathrm{A,\psi }}=a^{\mathrm{A}}+a^{\mathrm{A-\psi }}+a^{\mathrm{\psi }},
\label{3frs32}
\end{equation}%
where $a^{\mathrm{A}}$ contains only BRST generators from the
Abelian three-form sector, $a^{\mathrm{A-\psi }}$ describes the
cross-interactions between the two theories, and $a^{\mathrm{\psi
}}$ is responsible for the Rarita-Schwinger self-interactions. The
component $a^{\mathrm{\psi }}$ can be shown to take the same form
like in the case $D=4$ (see Ref.~\cite{boulcqg}) and satisfies
individually an equation of the type (\ref{3frs19}). It admits a
decomposition of the form
\begin{equation}
a^{\mathrm{\psi }}=a_{0}^{\mathrm{\psi }}+a_{1}^{\mathrm{\psi }},
\label{3frs33}
\end{equation}%
where
\begin{equation}
a_{1}^{\mathrm{\psi }}=\mathrm{i}m\psi _{\mu }^{\ast }\gamma ^{\mu }\xi
,\qquad a_{0}^{\mathrm{\psi }}=-\frac{9}{2}m\overline{\psi }_{\mu }\gamma
^{\mu \nu }\psi _{\nu },  \label{3frs34b}
\end{equation}%
with $m$ an arbitrary, real constant. Since $a^{\mathrm{A-\psi }}$ mixes the
variables from the three-form and the Rarita-Schwinger sectors and $a^{%
\mathrm{A}}$ depends only on the BRST generators from the three-form sector,
it follows that $a^{\mathrm{A-\psi }}$ and $a^{\mathrm{A}}$ are subject to
two separate equations
\begin{eqnarray}
sa^{\mathrm{A}} &=&\partial ^{\mu }m_{\mu }^{\mathrm{A}},  \label{3frs35a} \\
sa^{\mathrm{A-\psi }} &=&\partial ^{\mu }m_{\mu }^{\mathrm{A-\psi }}.
\label{3frs35b}
\end{eqnarray}%
The nontrivial solution $a^{\mathrm{A}}$ to (\ref{3frs35a}) has been
discussed in Ref.~\cite{SUGRAI} and reduces to
\begin{equation}
a^{\mathrm{A}}=q\varepsilon ^{\mu _{1}\ldots \mu _{11}}A_{\mu _{1}\mu
_{2}\mu _{3}}F_{\mu _{4}\ldots \mu _{7}}F_{\mu _{8}\ldots \mu _{11}},
\label{fo5}
\end{equation}%
with $q$ an arbitrary, real constant.

Let us analyze now the solutions to equation (\ref{3frs35b}). In agreement
with the previous results on $H^{\mathrm{inv}}\left( \delta |d\right) $, we
can always take the decomposition of $a^{\mathrm{A-\psi }}$ along the
antighost number to stop at antighost number equal to four
\begin{equation}
a^{\mathrm{A-\psi }}=a_{0}^{\mathrm{A-\psi }}+a_{1}^{\mathrm{A-\psi }%
}+a_{2}^{\mathrm{A-\psi }}+a_{3}^{\mathrm{A-\psi }}+a_{4}^{\mathrm{A-\psi }},
\label{3frs41}
\end{equation}
such that (\ref{3frs35b}) becomes equivalent with the tower of equations
\begin{eqnarray}
\gamma a_{4}^{\mathrm{A-\psi }} &=&0,  \label{3frs42a} \\
\delta a_{I}^{\mathrm{A-\psi }}+\gamma a_{I-1}^{\mathrm{A-\psi }}
&=&\partial _{\mu }m_{I-1}^{\mathrm{A-\psi }\mu },\qquad I=\overline{1,4}.
\label{3frs42b}
\end{eqnarray}
Recalling the results from the previous subsection on the cohomology $%
H(\gamma )$, it follows that the elements with pure ghost number four of a
basis in the space of polynomials in the ghosts $C$ and $\xi $ can be chosen
as
\begin{equation}
\left\{ \xi C,\left( \bar{\xi}\gamma _{\mu }\xi \right) \left( \bar{\xi}%
\gamma ^{\mu }\xi \right) ,\left( \bar{\xi}\gamma _{\mu \nu }\xi \right)
\left( \bar{\xi}\gamma ^{\mu \nu }\xi \right) ,\left( \bar{\xi}\gamma _{\mu
\nu \rho \lambda \sigma }\xi \right) \left( \bar{\xi}\gamma ^{\mu \nu \rho
\lambda \sigma }\xi \right) \right\} .  \label{3frs43}
\end{equation}
The solution to (\ref{3frs42a}) is obtained like in (\ref{3frs25}), by
`gluing' the general representative of $H^{\mathrm{inv}}\left( \delta
|d\right) $, namely $C^{*}$, to (\ref{3frs43})
\begin{eqnarray}
a_{4}^{\mathrm{A-\psi }} &=&v_{1}C^{*}\left( \bar{\xi}\gamma _{\mu }\xi
\right) \left( \bar{\xi}\gamma ^{\mu }\xi \right) +v_{2}C^{*}\left( \bar{\xi}%
\gamma _{\mu \nu }\xi \right) \left( \bar{\xi}\gamma ^{\mu \nu }\xi \right)
\notag \\
&&+v_{3}C^{*}\left( \bar{\xi}\gamma _{\mu \nu \rho \lambda \sigma }\xi
\right) \left( \bar{\xi}\gamma ^{\mu \nu \rho \lambda \sigma }\xi \right) ,
\label{3frs44}
\end{eqnarray}
where $\left( v_{i}\right) _{i=1,2,3}$ are some arbitrary constants [the
element $\xi C$ cannot be coupled to $C^{*}$ to form a Lorentz invariant
since $\xi $ is a Majorana spinor, so it is not eligible to enter (\ref%
{3frs44})]. Substituting (\ref{3frs44}) back in (\ref{3frs42b}) for $I=4$
and using definitions (\ref{3frs14a})--(\ref{3frs14h}), we obtain
\begin{eqnarray}
a_{3}^{\mathrm{A-\psi }} &=&-4C^{*\alpha }\left[ v_{1}\left( \bar{\xi}\gamma
_{\mu }\xi \right) \left( \bar{\xi}\gamma ^{\mu }\psi _{\alpha }\right)
+v_{2}\left( \bar{\xi}\gamma _{\mu \nu }\xi \right) \left( \bar{\xi}\gamma
^{\mu \nu }\psi _{\alpha }\right) \right.  \notag \\
&&\left. +v_{3}\left( \bar{\xi}\gamma _{\mu \nu \rho \lambda \sigma }\xi
\right) \left( \bar{\xi}\gamma ^{\mu \nu \rho \lambda \sigma }\psi _{\alpha
}\right) \right] .  \label{3frs45}
\end{eqnarray}
By applying the Koszul-Tate differential on (\ref{3frs45}), we find
\begin{eqnarray}
\delta a_{3}^{\mathrm{A-\psi }} &=&\gamma \left\{ 8C^{*\alpha \beta }\left[
v_{1}\left( \bar{\xi}\gamma _{\mu }\psi _{\alpha }\right) \left( \bar{\xi}%
\gamma ^{\mu }\psi _{\beta }\right) \right. \right.  \notag \\
&&\left. +v_{2}\left( \bar{\xi}\gamma _{\mu \nu }\psi _{\alpha }\right)
\left( \bar{\xi}\gamma ^{\mu \nu }\psi _{\beta }\right) +v_{3}\left( \bar{\xi%
}\gamma _{\mu \nu \rho \lambda \sigma }\psi _{\alpha }\right) \left( \bar{\xi%
}\gamma ^{\mu \nu \rho \lambda \sigma }\psi _{\beta }\right) \right]  \notag
\\
&&-4C^{*\alpha \beta }\left[ v_{1}\left( \bar{\xi}\gamma _{\mu }\xi \right)
\left( \bar{\psi}_{\alpha }\gamma ^{\mu }\psi _{\beta }\right) \right.
\notag \\
&&\left. \left. +v_{2}\left( \bar{\xi}\gamma _{\mu \nu }\xi \right) \left(
\psi _{\alpha }\gamma ^{\mu \nu }\psi _{\beta }\right) +v_{3}\left( \bar{\xi}%
\gamma _{\mu \nu \rho \lambda \sigma }\xi \right) \left( \bar{\psi}_{\alpha
}\gamma ^{\mu \nu \rho \lambda \sigma }\psi _{\beta }\right) \right] \right\}
\notag \\
&&-4C^{*\alpha \beta }\left[ v_{1}\left( \bar{\xi}\gamma _{\mu }\xi \right)
\left( \bar{\xi}\gamma ^{\mu }\partial _{[\alpha }\psi _{\beta ]}\right)
+v_{2}\left( \bar{\xi}\gamma _{\mu \nu }\xi \right) \left( \bar{\xi}\gamma
^{\mu \nu }\partial _{[\alpha }\psi _{\beta ]}\right) \right.  \notag \\
&&\left. +v_{3}\left( \bar{\xi}\gamma _{\mu \nu \rho \lambda \sigma }\xi
\right) \left( \bar{\xi}\gamma ^{\mu \nu \rho \lambda \sigma }\partial
_{[\alpha }\psi _{\beta ]}\right) \right] +\partial _{\mu }m_{2}^{\mathrm{%
A-\psi \;}\mu }.  \label{3frs46}
\end{eqnarray}
Comparing (\ref{3frs46}) with (\ref{3frs35b}) for $I=3$ it follows that $%
a_{3}^{\mathrm{A-\psi }}$ provides a consistent $a_{2}^{\mathrm{A-\psi }}$
if the quantity
\begin{eqnarray}
\pi &=&-4C^{*\alpha \beta }\left[ v_{1}\left( \bar{\xi}\gamma _{\mu }\xi
\right) \left( \bar{\xi}\gamma ^{\mu }\partial _{[\alpha }\psi _{\beta
]}\right) +v_{2}\left( \bar{\xi}\gamma _{\mu \nu }\xi \right) \left( \bar{\xi%
}\gamma ^{\mu \nu }\partial _{[\alpha }\psi _{\beta ]}\right) \right.  \notag
\\
&&\left. +v_{3}\left( \bar{\xi}\gamma _{\mu \nu \rho \lambda \sigma }\xi
\right) \left( \bar{\xi}\gamma ^{\mu \nu \rho \lambda \sigma }\partial
_{[\alpha }\psi _{\beta ]}\right) \right] ,  \label{zzz1}
\end{eqnarray}
can be written in a $\gamma $-exact modulo $d$ form
\begin{equation}
\pi =\gamma w+\partial _{\mu }\theta ^{\mu }.  \label{zzz2}
\end{equation}
Assume that (\ref{zzz2}) holds. Taking its Euler-Lagrange (EL) derivatives
with respect to $C^{*\alpha \beta }$ we get
\begin{equation}
\frac{\delta ^{\mathrm{L}}\pi }{\delta C^{*\alpha \beta }}=\gamma \left(
\frac{\delta ^{\mathrm{L}}w}{\delta C^{*\alpha \beta }}\right) .
\label{zzz3}
\end{equation}
On the other hand, from (\ref{zzz1}) by direct computation we infer
\begin{eqnarray}
\frac{\delta ^{\mathrm{L}}\pi }{\delta C^{*\alpha \beta }} &=&-4\left[
v_{1}\left( \bar{\xi}\gamma _{\mu }\xi \right) \left( \bar{\xi}\gamma ^{\mu
}\partial _{[\alpha }\psi _{\beta ]}\right) +v_{2}\left( \bar{\xi}\gamma
_{\mu \nu }\xi \right) \left( \bar{\xi}\gamma ^{\mu \nu }\partial _{[\alpha
}\psi _{\beta ]}\right) \right.  \notag \\
&&\left. +v_{3}\left( \bar{\xi}\gamma _{\mu \nu \rho \lambda \sigma }\xi
\right) \left( \bar{\xi}\gamma ^{\mu \nu \rho \lambda \sigma }\partial
_{[\alpha }\psi _{\beta ]}\right) \right] .  \label{zzz4}
\end{eqnarray}
Thus, equation (\ref{zzz3}) restricts $\delta ^{\mathrm{L}}\pi /\delta
C^{*\alpha \beta }$ to be a trivial element of $H(\gamma )$, while (\ref%
{zzz4}) emphasizes that $\delta ^{\mathrm{L}}\pi /\delta C^{*\alpha \beta }$
is a nontrivial element from $H(\gamma )$ (because each term from the
right-hand side of (\ref{zzz4}) is so), such that the only possibility is
that $\delta ^{\mathrm{L}}\pi /\delta C^{*\alpha \beta }$ must vanish
\begin{equation}
\frac{\delta ^{\mathrm{L}}\pi }{\delta C^{*\alpha \beta }}=0.  \label{mmn2}
\end{equation}
This further implies, by means of (\ref{zzz4}), that we must set zero all
the constants that parameterize $a_{4}^{\mathrm{A-\psi }}$
\begin{equation}
v_{1}=v_{2}=v_{3}=0,  \label{3frs47}
\end{equation}
so in the end we have that
\begin{equation}
a_{4}^{\mathrm{A-\psi }}=a_{3}^{\mathrm{A-\psi }}=0.  \label{3frs48}
\end{equation}

As a consequence, decomposition (\ref{3frs41}) can stop earliest at
antighost number three, $a^{\mathrm{A-\psi }}=a_{0}^{\mathrm{A-\psi }%
}+a_{1}^{\mathrm{A-\psi }}+a_{2}^{\mathrm{A-\psi }}+a_{3}^{\mathrm{A-\psi }}$%
, where $a_{3}^{\mathrm{A-\psi }}$ satisfies the equation $\gamma a_{3}^{%
\mathrm{A-\psi }}=0$. According to (\ref{3frs25}), (\ref{3frs30b}) and
recalling the assumption that $a_{3}^{\mathrm{A-\psi }}$ mixes the BRST
generators of the three-form with those from the Rarita-Schwinger sector, it
results that the solution to this equation reads as $a_{3}^{\mathrm{A-\psi }%
}=C_{\mu }^{\ast }e^{\mu }\left( \xi \right) $, where $e^{\mu }\left( \xi
\right) $ denote the vector-like elements of pure ghost number three of a
basis in the space of polynomials in the ghost $\xi $. Since $\mathrm{pgh}%
\left( \xi \right) =1$, it follows that $e^{\mu }\left( \xi \right) $
necessarily contains three spinors of the type $\xi $ and therefore we can
set $a_{3}^{\mathrm{A-\psi }}=0$ because one cannot construct a Lorentz
eleven-dimensional vector out of three spinors.

Thus, we can write
\begin{equation}
a^{\mathrm{A-\psi }}=a_{0}^{\mathrm{A-\psi }}+a_{1}^{\mathrm{A-\psi }%
}+a_{2}^{\mathrm{A-\psi }},  \label{aintagh2}
\end{equation}%
such that equation(\ref{3frs35b}) becomes equivalent to
\begin{eqnarray}
\gamma a_{2}^{\mathrm{A-\psi }} &=&0,  \label{3frs49a} \\
\delta a_{I}^{\mathrm{A-\psi }}+\gamma a_{I-1}^{\mathrm{A-\psi }}
&=&\partial _{\mu }m_{I-1}^{\mathrm{A-\psi \;}\mu },\qquad I=\overline{1,2}.
\label{3frs49b}
\end{eqnarray}%
Because the elements of pure ghost number two of a basis in the space of
polynomials in the ghost $\xi $ read as
\begin{equation}
\left\{ \left( \bar{\xi}\gamma _{\mu }\xi \right) ,\left( \bar{\xi}\gamma
_{\mu \nu }\xi \right) ,\left( \bar{\xi}\gamma _{\mu \nu \rho \lambda \sigma
}\xi \right) \right\}  \label{3frs50}
\end{equation}%
(the ghost for ghost for ghost $C$ is not eligible as $\mathrm{pgh}\left(
C\right) =3$) and the representatives of $H_{2}^{\mathrm{inv}}\left( \delta
|d\right) $ are given by (\ref{3frs30c}), we observe that the only
combination that might generate cross-interactions remains
\begin{equation}
a_{2}^{\mathrm{A-\psi }}=\frac{\tilde{k}}{2}C^{\ast \mu \nu }\bar{\xi}\gamma
_{\mu \nu }\xi ,  \label{3frs51}
\end{equation}%
where $\tilde{k}$ is an arbitrary constant. Replacing (\ref{3frs51}) in (\ref%
{3frs49b}) for $I=2$ we determine $a_{1}^{\mathrm{A-\psi }}$ under the form
\begin{equation}
a_{1}^{\mathrm{A-\psi }}=-3\tilde{k}A^{\ast \mu \nu \rho }\bar{\xi}\gamma
_{\mu \nu }\psi _{\rho }+\bar{a}_{1}^{\mathrm{A-\psi }},  \label{3frs52}
\end{equation}%
where $\bar{a}_{1}^{\mathrm{A-\psi }}$ is the general solution to the
`homogeneous' equation
\begin{equation}
\gamma \bar{a}_{1}^{\mathrm{A-\psi }}=0.  \label{3frs53}
\end{equation}%
It is expressed by
\begin{equation}
\bar{a}_{1}^{\mathrm{A-\psi }}=\left( \psi ^{\ast \mu }M_{\mu }+A^{\ast \mu
\nu \rho }N_{\mu \nu \rho }\right) \xi ,  \label{3frs54}
\end{equation}%
with $N_{\mu \nu \rho }$ the components of a real, fermionic,
gauge-invariant, completely antisymmetric spinor tensor and $M_{\mu }$ some
bosonic, gauge-invariant, $11\times 11$ matrices, which in addition must
explicitly depend on the three-form field strength $F_{\mu \nu \rho \lambda
} $ in order to provide cross-interactions. By applying $\delta $ on (\ref%
{3frs52}) with $\bar{a}_{1}^{\mathrm{A-\psi }}$ of the form (\ref{3frs54}),
we obtain
\begin{equation}
\delta a_{1}^{\mathrm{A-\psi }}=\gamma d_{0}+e_{0}+\partial _{\mu }m^{\mu },
\label{dpea1}
\end{equation}%
where
\begin{eqnarray}
d_{0} &=&\left( \frac{1}{4}\tilde{k}\bar{\psi}_{\mu }\gamma _{\nu \rho }\psi
_{\lambda }+\frac{1}{3!}N_{\mu \nu \rho }\psi _{\lambda }\right) F^{\mu \nu
\rho \lambda },  \label{b0} \\
e_{0} &=&\frac{1}{4}\tilde{k}\bar{\xi}\gamma _{\mu \nu }\left( \partial
_{\rho }\psi _{\lambda }\right) F^{\mu \nu \rho \lambda }+\mathrm{i}\bar{\psi%
}_{\lambda }\gamma ^{\rho \lambda \mu }\left( \partial _{\rho }M_{\mu
}\right) \xi  \notag \\
&&+\mathrm{i}\bar{\psi}_{\lambda }\gamma ^{\rho \lambda \mu }M_{\mu
}\partial _{\rho }\xi +\frac{1}{3!}\left( \partial _{\lambda }N_{\mu \nu
\rho }\right) \xi F^{\mu \nu \rho \lambda }.  \label{c0}
\end{eqnarray}%
The condition that (\ref{dpea1}) is expressed like in (\ref{3frs49b}) for $%
I=1$ restricts $e_{0}$ expressed by (\ref{c0}) to be $\gamma $-exact modulo $%
d$%
\begin{equation}
e_{0}=\gamma p+\partial _{\mu }n^{\mu }.  \label{condc0}
\end{equation}%
Recalling the requirement that the quantities $N_{\mu \nu \rho }$ are
spinor-like and gauge-invariant, we deduce that the most general
representation of these elements is
\begin{equation}
N_{\mu \nu \rho }=\partial _{\lbrack \alpha }\bar{\psi}_{\beta ]}N_{\mu \nu
\rho }^{\;\;\;\;\alpha \beta },  \label{3frs55}
\end{equation}%
where $N_{\mu \nu \rho }^{\;\;\;\;\alpha \beta }$ are also gauge-invariant.
As $c_{0}$ from (\ref{c0}) involves terms with different numbers of
derivatives, it is useful to decompose the functions $M_{\mu }$ and $N_{\mu
\nu \rho }^{\;\;\;\;\alpha \beta }$ according to the number of spacetime
derivatives
\begin{eqnarray}
M_{\mu } &=&\overset{\left( 1\right) }{M}_{\mu }+\overset{\left( 2\right) }{M%
}_{\mu }+\overset{\left( 3\right) }{M}_{\mu }+\cdots ,  \label{dez1} \\
N_{\mu \nu \rho }^{\;\;\;\;\alpha \beta } &=&\overset{\left( 0\right) }{N}%
_{\mu \nu \rho }^{\;\;\;\;\alpha \beta }+\overset{\left( 1\right) }{N}_{\mu
\nu \rho }^{\;\;\;\;\alpha \beta }+\cdots ,  \label{dez2}
\end{eqnarray}%
where $\overset{\left( k\right) }{M}_{\mu }$ and $\overset{\left( k\right) }{%
N}_{\mu \nu \rho }^{\;\;\;\;\alpha \beta }$contain precisely $k$ derivatives
[(\ref{dez1}) cannot contain a derivative-free term because, as we have
emphasized before, $M_{\mu }$ depends at least linearly on $F^{\mu \nu \rho
\lambda }$]. Inserting (\ref{dez1}) and (\ref{dez2}) in (\ref{c0}) and
projecting (\ref{condc0}) on the various numbers of derivatives, we find the
equivalent tower of equations
\begin{eqnarray}
&&\frac{1}{4}\tilde{k}\bar{\xi}\gamma _{\mu \nu }\left( \partial _{\rho
}\psi _{\lambda }\right) F^{\mu \nu \rho \lambda }+\mathrm{i}\bar{\psi}%
_{\lambda }\gamma ^{\rho \lambda \mu }\left( \partial _{\rho }\overset{%
\left( 1\right) }{M}_{\mu }\right) \xi  \notag \\
&&+\mathrm{i}\bar{\psi}_{\lambda }\gamma ^{\rho \lambda \mu }\overset{\left(
1\right) }{M}_{\mu }\partial _{\rho }\xi =\gamma \overset{\left( 1\right) }{p%
}+\partial _{\mu }\overset{\left( 1\right) }{n}^{\mu },  \label{ec1} \\
&&\mathrm{i}\bar{\psi}_{\lambda }\gamma ^{\rho \lambda \mu }\left( \partial
_{\rho }\overset{\left( 2\right) }{M}_{\mu }\right) \xi +\mathrm{i}\bar{\psi}%
_{\lambda }\gamma ^{\rho \lambda \mu }\overset{\left( 2\right) }{M}_{\mu
}\partial _{\rho }\xi  \notag \\
&&+\frac{1}{3!}\partial _{\lambda }\left( \partial _{\lbrack \alpha }\bar{%
\psi}_{\beta ]}\overset{\left( 0\right) }{N}_{\mu \nu \rho }^{\;\;\;\;\alpha
\beta }\right) \xi F^{\mu \nu \rho \lambda }=\gamma \overset{\left( 2\right)
}{p}+\partial _{\mu }\overset{\left( 2\right) }{n}^{\mu },  \label{eck} \\
&&\mathrm{i}\bar{\psi}_{\lambda }\gamma ^{\rho \lambda \mu }\left( \partial
_{\rho }\overset{\left( k\right) }{M}_{\mu }\right) \xi +\mathrm{i}\bar{\psi}%
_{\lambda }\gamma ^{\rho \lambda \mu }\overset{\left( k\right) }{M}_{\mu
}\partial _{\rho }\xi  \notag \\
&&+\frac{1}{3!}\partial _{\lambda }\left( \partial _{\lbrack \alpha }\bar{%
\psi}_{\beta ]}\overset{\left( k-2\right) }{N}_{\mu \nu \rho
}^{\;\;\;\;\alpha \beta }\right) \xi F^{\mu \nu \rho \lambda }=\gamma
\overset{\left( k\right) }{p}+\partial _{\mu }\overset{\left( k\right) }{n}%
^{\mu },\qquad k\geq 3.  \label{ecz1}
\end{eqnarray}%
Equations (\ref{ecz1}) would lead to interaction vertices with more than two
spacetime derivatives, so, in agreement with our hypothesis on the
conservation of the number of derivatives on each field with respect to the
free theory, they must be discarded
\begin{eqnarray}
\overset{\left( k\right) }{M}_{\mu } &=&0,\qquad k\geq 3,  \label{zq1} \\
\overset{\left( k\right) }{N}_{\mu \nu \rho }^{\;\;\;\;\alpha \beta }
&=&0,\qquad k\geq 1,  \label{zq2}
\end{eqnarray}%
which ensures $\overset{\left( k\right) }{p}=0$ for $k\geq 3$ in (\ref%
{condc0}). As the matrices $\overset{\left( 1\right) }{M}_{\mu }$ are linear
in the three-form field strength, they can be generally represented in the
form
\begin{equation}
\overset{\left( 1\right) }{M}_{\mu }=k_{1}F_{\mu }^{\;\;\alpha \beta
\gamma }\gamma _{\alpha \beta \gamma }+k_{2}F^{\alpha \beta \gamma
\delta }\gamma _{\mu \alpha \beta \gamma \delta  }, \label{reprezM1}
\end{equation}%
with $k_{1}$ and $k_{2}$ some arbitrary constants. Based on (\ref{reprezM1}%
), the left-hand side of (\ref{ec1}) becomes
\begin{eqnarray}
&&\frac{1}{4}\tilde{k}\bar{\xi}\gamma _{\mu \nu }\left( \partial _{\rho
}\psi _{\lambda }\right) F^{\mu \nu \rho \lambda }+\mathrm{i}\bar{\psi}%
_{\lambda }\gamma ^{\rho \lambda \mu }\left( \partial _{\rho }\overset{%
\left( 1\right) }{M}_{\mu }\right) \xi  \notag \\
&&+\mathrm{i}\bar{\psi}_{\lambda }\gamma ^{\rho \lambda \mu }\overset{\left(
1\right) }{M}_{\mu }\gamma \psi _{\rho }=\gamma \left[ \left( -\frac{3}{8}%
\left( k_{1}+8k_{2}\right) \mathrm{i}\bar{\psi}^{\alpha }\gamma _{\mu \nu
\rho \lambda }\psi _{\alpha }\right. \right.  \notag \\
&&\left. \left. -\frac{1}{2}\left( k_{1}+5k_{2}\right) \mathrm{i}\bar{\psi}%
^{\alpha }\gamma _{\alpha \beta \mu \nu \rho \lambda }\psi ^{\beta }\right)
F^{\mu \nu \rho \lambda }\right]  \notag \\
&&+\frac{1}{2}\left( \tilde{k}-2\cdot 3!\mathrm{i}k_{1}-7\cdot 4!\mathrm{i}%
k_{2}\right) \bar{\xi}\gamma _{\mu \nu }\left( \partial _{\rho }\psi
_{\lambda }\right) F^{\mu \nu \rho \lambda }  \notag \\
&&+\left( k_{1}+8k_{2}\right) \left[ -\frac{3}{4}\mathrm{i}\left(
\partial ^{\alpha }\bar{\psi}_{\alpha }\right) \gamma _{\mu \nu \rho
\lambda }\xi
\right.  \notag \\
&&\left. +3\mathrm{i}\left( \partial _{\mu }\bar{\psi}^{\alpha }\right)
\gamma _{\alpha \nu \rho \lambda }\xi +3\mathrm{i}\bar{\psi}_{\mu }\gamma
_{\alpha \nu \rho \lambda }\left( \gamma \psi ^{\alpha }\right) \right]
F^{\mu \nu \rho \lambda }+\partial _{\mu }\overset{\left( 1\right) }{n}^{\mu
}.  \label{ec1.1}
\end{eqnarray}%
Asking now that the right-hand side of (\ref{ec1.1}) satisfies (\ref{ec1}),
we find the restrictions
\begin{equation}
k_{1}=\frac{1}{9}\mathrm{i}\tilde{k},\qquad k_{2}=-\frac{1}{3\cdot 4!}%
\mathrm{i}\tilde{k},  \label{3frs61}
\end{equation}%
which further produce
\begin{eqnarray}
\overset{\left( 1\right) }{M}_{\mu } &=&\mathrm{i}\tilde{k}\left( \frac{1}{9}%
F_{\mu }^{\;\;\alpha \beta \gamma }\gamma _{\alpha \beta \gamma }-\frac{1}{%
3\cdot 4!}F^{\alpha \beta \gamma \delta }\gamma _{\mu \alpha \beta \gamma
\delta }\right) ,  \label{bb1} \\
\overset{\left( 1\right) }{p} &=&\frac{1}{2\cdot 4!}\tilde{k}\bar{\psi}%
^{\alpha }\gamma _{\alpha \beta \mu \nu \rho \lambda }\psi ^{\beta }F^{\mu
\nu \rho \lambda }.  \label{bb2}
\end{eqnarray}%
Next, we approach equation (\ref{eck}). Due to the fact that each $\overset{%
\left( 2\right) }{M}_{\mu }$ is a gauge-invariant, $11\times 11$ matrix with
two spacetime derivatives, it contains precisely two three-form field
strengths (since it cannot depend on $\partial _{\lbrack \alpha }\bar{\psi}%
_{\beta ]}$, which is a spinor). As the elements $\overset{\left( 0\right) }{%
N}_{\mu \nu \rho }^{\;\;\;\;\alpha \beta }$ are derivative-free and
gauge-invariant, they can only be constant. Based on the last two
observations, we observe that each of the first two terms from the left-hand
side of equation (\ref{eck}) comprises two three-form field strengths, while
the last term is only linear in $F^{\mu \nu \rho \lambda }$, such that (\ref%
{eck}) splits into two separate equations
\begin{eqnarray}
\mathrm{i}\bar{\psi}_{\lambda }\gamma ^{\rho \lambda \mu }\left( \partial
_{\rho }\overset{\left( 2\right) }{M}_{\mu }\right) \xi +\mathrm{i}\bar{\psi}%
_{\lambda }\gamma ^{\rho \lambda \mu }\overset{\left( 2\right) }{M}_{\mu
}\partial _{\rho }\xi &=&\gamma \overset{\left( 2\right) }{p_{1}}+\partial
_{\mu }\overset{\left( 2\right) }{n_{1}}^{\mu },  \label{ec2.1} \\
\frac{1}{3!}\partial _{\lambda }\left( \partial _{\lbrack \alpha }\bar{\psi}%
_{\beta ]}\overset{\left( 0\right) }{N}_{\mu \nu \rho }^{\;\;\;\;\alpha
\beta }\right) \xi F^{\mu \nu \rho \lambda } &=&\gamma \overset{\left(
2\right) }{p_{2}}+\partial _{\mu }\overset{\left( 2\right) }{n_{2}}^{\mu }.
\label{ec2.2}
\end{eqnarray}%
The left-hand side of (\ref{ec2.1}) is $\gamma $-exact modulo $d$ if the
following conditions are simultaneously satisfied
\begin{eqnarray}
\gamma _{0}\gamma ^{\rho \lambda \mu }\overset{\left( 2\right) }{M}_{\mu }
&=&-\left( \gamma _{0}\gamma ^{\lambda \rho \mu }\overset{\left( 2\right) }{M%
}_{\mu }\right) ^{T},  \label{cond1} \\
\partial _{\lbrack \rho }\overset{\left( 2\right) }{M}_{\mu ]} &=&0.
\label{cond2}
\end{eqnarray}%
In order to investigate the former condition, we represent $\overset{\left(
2\right) }{M}_{\mu }$ in terms of a basis in the space of $\gamma $-matrices
\begin{eqnarray}
\overset{\left( 2\right) }{M}_{\mu } &=&\overset{\left( 2\right) }{\bar{M}}%
_{\mu }\mathbf{1}+\overset{\left( 2\right) }{\bar{M}}_{\mu }^{\alpha }\gamma
_{\alpha }+\overset{\left( 2\right) }{\bar{M}}_{\mu }^{\alpha \beta }\gamma
_{\alpha \beta }+\overset{\left( 2\right) }{\bar{M}}_{\mu }^{\alpha \beta
\gamma }\gamma _{\alpha \beta \gamma }  \notag \\
&&+\overset{\left( 2\right) }{\bar{M}}_{\mu }^{\alpha \beta \gamma \delta
}\gamma _{\alpha \beta \gamma \delta }+\overset{\left( 2\right) }{\bar{M}}%
_{\mu }^{\alpha \beta \gamma \delta \varepsilon }\gamma _{\alpha \beta
\gamma \delta \varepsilon },  \label{reprezM2}
\end{eqnarray}%
where each of the coefficients $\overset{\left( 2\right) }{\bar{M}}_{\mu }$,%
\textbf{\ }$\overset{\left( 2\right) }{\bar{M}}_{\mu }^{\alpha }$, $\overset{%
\left( 2\right) }{\bar{M}}_{\mu }^{\alpha \beta }$, $\overset{\left(
2\right) }{\bar{M}}_{\mu }^{\alpha \beta \gamma }$, $\overset{\left(
2\right) }{\bar{M}}_{\mu }^{\alpha \beta \gamma \delta }$, and $\overset{%
\left( 2\right) }{\bar{M}}_{\mu }^{\alpha \beta \gamma \delta \varepsilon }$
is a function with precisely two three-form field strengths. This dependence
implies the vanishing of all coefficients with an odd number of indices
\begin{equation}
\overset{\left( 2\right) }{\bar{M}}_{\mu }=0,\qquad \overset{\left( 2\right)
}{\bar{M}}_{\mu }^{\alpha \beta }=0,\qquad \overset{\left( 2\right) }{\bar{M}%
}_{\mu }^{\alpha \beta \gamma \delta }=0.  \label{megal0}
\end{equation}%
Inserting (\ref{megal0}) in (\ref{reprezM2}) we find by direct computation
the relation
\begin{eqnarray}
&\gamma ^{\rho \lambda \mu }\overset{\left( 2\right) }{M}_{\mu }=&\overset{%
\left( 2\right) }{\bar{M}}_{\mu }^{\alpha }\left( \delta _{\alpha }^{[\mu
}\gamma _{\left. {}\right. }^{\rho \lambda ]}+\gamma _{\qquad \alpha }^{\rho
\lambda \mu }\right) +\overset{\left( 2\right) }{\bar{M}}_{\mu }^{\alpha
\beta \gamma }\left( \delta _{\lbrack \alpha }^{\mu }\delta _{\beta
}^{\lambda }\delta _{\gamma ]}^{\rho }\mathbf{1}+\delta _{\lbrack \alpha
}^{[\mu }\delta _{\beta }^{\lambda }\gamma _{\ \ \gamma ]}^{\rho ]}\right.
\notag \\
&&\left. +\delta _{\lbrack \alpha }^{[\mu }\gamma _{\qquad \beta \gamma
]}^{\rho \lambda ]}+\gamma _{\qquad \alpha \beta \gamma }^{\rho \lambda \mu
}\right) +\overset{\left( 2\right) }{\bar{M}}_{\mu }^{\alpha \beta \gamma
\delta \varepsilon }\left( \delta _{\lbrack \alpha }^{\mu }\delta _{\beta
}^{\lambda }\delta _{\gamma }^{\rho }\gamma _{\delta \varepsilon ]}^{\left.
{}\right. }\right.  \notag \\
&&\left. +\delta _{\lbrack \alpha }^{[\mu }\delta _{\beta }^{\lambda }\gamma
_{\ \ \gamma \delta \varepsilon ]}^{\rho ]}+\delta _{\lbrack \alpha }^{[\mu
}\gamma _{\qquad \beta \gamma \delta \varepsilon ]}^{\rho \lambda ]}+\gamma
_{\qquad \alpha \beta \gamma \delta \varepsilon }^{\rho \lambda \mu }\right)
.  \label{gamaM2}
\end{eqnarray}%
Looking at (\ref{gamaM2}), we remark that the terms $3!\overset{\left(
2\right) }{\bar{M}}_{\mu }^{\mu \lambda \rho }\mathbf{1}$ and $\overset{%
\left( 2\right) }{\bar{M}}_{\mu }^{\alpha \beta \gamma \delta \varepsilon
}\gamma _{\qquad \alpha \beta \gamma \delta \varepsilon }^{\rho \lambda \mu
} $ appearing in $\gamma ^{\rho \lambda \mu }\overset{\left( 2\right) }{M}%
_{\mu }$ break condition (\ref{cond1}), so we must set
\begin{equation}
\overset{\left( 2\right) }{\bar{M}}_{\mu }^{\alpha \beta \gamma }=0,\qquad
\overset{\left( 2\right) }{\bar{M}}_{\mu }^{\alpha \beta \gamma \delta
\varepsilon }=0.  \label{megal0.1}
\end{equation}%
The last result replaced in (\ref{gamaM2}) yields
\begin{equation}
\gamma ^{\rho \lambda \mu }\overset{\left( 2\right) }{M}_{\mu }=\overset{%
\left( 2\right) }{\bar{M}}_{\mu }^{\alpha }\left( \delta _{\alpha }^{[\mu
}\gamma _{\left. {}\right. }^{\rho \lambda ]}+\gamma _{\qquad \alpha }^{\rho
\lambda \mu }\right) .  \label{gamaM2.1}
\end{equation}%
It is clear that $\overset{\left( 2\right) }{\bar{M}}_{\mu }^{\alpha }\gamma
_{\qquad \alpha }^{\rho \lambda \mu }$ from (\ref{gamaM2.1}) cannot fulfill (%
\ref{cond1}), so we must take
\begin{equation}
\overset{\left( 2\right) }{\bar{M}}_{\mu }^{\alpha }=0,  \label{megal0.2}
\end{equation}%
which, together with (\ref{megal0}), (\ref{megal0.1}), and (\ref{megal0.2}),
lead to the result
\begin{equation}
\overset{\left( 2\right) }{M}_{\mu }=0,  \label{M2egalzero}
\end{equation}%
so we can only have $\overset{\left( 2\right) }{p_{1}}=0$ in (\ref{ec2.1}).
Now, we investigate equation (\ref{ec2.2}). Direct computation provides
\begin{eqnarray}
\frac{1}{3!}\partial _{\lambda }\left( \partial _{\lbrack \alpha }\bar{\psi}%
_{\beta ]}\overset{\left( 0\right) }{N}_{\mu \nu \rho }^{\;\;\;\;\alpha
\beta }\right) \xi F^{\mu \nu \rho \lambda } &=&\gamma \left( -\frac{1}{3!}%
\partial _{\lbrack \alpha }\bar{\psi}_{\beta ]}\overset{\left( 0\right) }{N}%
_{\mu \nu \rho }^{\;\;\;\;\alpha \beta }\psi _{\lambda }F^{\mu \nu \rho
\lambda }\right)  \notag \\
&&-\frac{1}{3!}\partial _{\lbrack \alpha }\bar{\psi}_{\beta ]}\overset{%
\left( 0\right) }{N}_{\mu \nu \rho }^{\;\;\;\;\alpha \beta }\xi \partial
_{\lambda }F^{\mu \nu \rho \lambda }+\partial _{\mu }s^{\mu }.  \label{zx1}
\end{eqnarray}%
Assume that the second term from the right-hand side of (\ref{zx1}) would
give a $\gamma $-exact modulo $d$ quantity. Comparing (\ref{zx1}) to (\ref%
{eck}), we find that
\begin{equation}
\overset{\left( 2\right) }{p_{2}}=-\frac{1}{3!}\partial _{\lbrack \alpha }%
\bar{\psi}_{\beta ]}\overset{\left( 0\right) }{N}_{\mu \nu \rho
}^{\;\;\;\;\alpha \beta }\psi _{\lambda }F^{\mu \nu \rho \lambda }+\cdots .
\label{p22a}
\end{equation}%
It is simple to see that $\overset{\left( 2\right) }{p_{2}}$ (which
contributes to $a_{0}^{\mathrm{A-\psi }}$) produces field equations for the
Rarita-Schwinger field with two spacetime derivatives, which disagrees with
requirement (i) from the beginning of this section related to the derivative
order assumption. Thus, we must set
\begin{equation}
\overset{\left( 0\right) }{N}_{\mu \nu \rho }^{\;\;\;\;\alpha \beta }=0,
\label{Negal0}
\end{equation}%
which yields $\overset{\left( 2\right) }{p_{2}}=0$.

Inserting (\ref{zq1}), (\ref{bb1}), and (\ref{M2egalzero}) into (\ref{dez1})
and respectively (\ref{zq2}) and (\ref{Negal0}) into (\ref{dez2}), and then
substituting the resulting expressions of (\ref{dez1}) and (\ref{dez2}) in (%
\ref{3frs54}), we obtain the general form of the solution $\bar{a}_{1}^{%
\mathrm{A-\psi }}$, such that (\ref{3frs52}) takes the final form
\begin{equation}
a_{1}^{\mathrm{A-\psi }}=-3\tilde{k}A^{\ast \mu \nu \rho }\bar{\xi}\gamma
_{\mu \nu }\psi _{\rho }+\mathrm{i}\tilde{k}\psi ^{\ast \mu }\left( \frac{1}{%
9}F_{\mu \nu \rho \lambda }\gamma ^{\nu \rho \lambda }-\frac{1}{3\cdot 4!}%
F^{\nu \rho \lambda \sigma }\gamma _{\mu \nu \rho \lambda \sigma }\right)
\xi .  \label{wq1}
\end{equation}%
Accordingly, we find that $a_{0}^{\mathrm{A-\psi }}$ as solution to equation
(\ref{3frs49b}) for $I=1$ reads as
\begin{equation}
a_{0}^{\mathrm{A-\psi }}=-\frac{1}{4}\tilde{k}\bar{\psi}_{\mu }\gamma _{\nu
\rho }\psi _{\lambda }F^{\mu \nu \rho \lambda }-\frac{1}{2\cdot 4!}\tilde{k}%
\bar{\psi}^{\alpha }\gamma _{\alpha \beta \mu \nu \rho \lambda }\psi ^{\beta
}F^{\mu \nu \rho \lambda }.  \label{3frs62}
\end{equation}%
Replacing now (\ref{3frs51}) and (\ref{wq1})--(\ref{3frs62}) into (\ref%
{aintagh2}), we find that the interacting part of the first-order
deformation of the solution to the master equation becomes
\begin{eqnarray}
S_{1}^{\mathrm{A-\psi }} &=&\int d^{11}x\left( a_{2}^{\mathrm{A-\psi }%
}+a_{1}^{\mathrm{A-\psi }}+a_{0}^{\mathrm{A-\psi }}\right)  \notag \\
&\equiv &\tilde{k}\int d^{11}x\left( \frac{1}{2}C^{\ast \mu \nu }\bar{\xi}%
\gamma _{\mu \nu }\xi -3A^{\ast \mu \nu \rho }\bar{\xi}\gamma _{\mu \nu
}\psi _{\rho }\right.  \notag \\
&&+\frac{\mathrm{i}}{9}\psi ^{\ast \mu }F_{\mu \nu \rho \lambda
}\gamma ^{\nu \rho \lambda }\xi -\frac{\mathrm{i}}{3\cdot 4!}\psi
^{\ast \mu }F^{\nu \rho \lambda \sigma
}\gamma _{\mu \nu \rho \lambda \sigma }\xi  \notag \\
&&\left. -\frac{1}{4}\bar{\psi}_{\mu }\gamma _{\nu \rho }\psi _{\lambda
}F^{\mu \nu \rho \lambda }-\frac{1}{2\cdot 4!}\bar{\psi}^{\alpha }\gamma
_{\alpha \beta \mu \nu \rho \lambda }\psi ^{\beta }F^{\mu \nu \rho \lambda
}\right) .  \label{3frs63}
\end{eqnarray}%
In what follows we employ the notation
\begin{equation}
S_{1}^{\prime \mathrm{A-\psi }}=S_{1}^{\mathrm{A-\psi }}+\int d^{11}x\,a^{%
\mathrm{A}},  \label{qw4}
\end{equation}%
with $a^{\mathrm{A}}$ given by (\ref{fo5}), so the complete expression of
the first-order deformation of the solution to the master equation for the
model under consideration is (see (\ref{3frs32}))
\begin{equation}
S_{1}^{\mathrm{A,\psi }}=S_{1}^{\prime \mathrm{A-\psi }}+S_{1}^{\mathrm{\psi
}},  \label{def1}
\end{equation}%
where $S_{1}^{\mathrm{\psi }}$ is the component corresponding to the
Rarita-Schwinger sector
\begin{equation}
S_{1}^{\mathrm{\psi }}=\int d^{11}xa^{\mathrm{\psi }},  \label{3frsdefrs}
\end{equation}%
and the integrand $a^{\mathrm{\psi }}$ can be read from (\ref{3frs33}) and (%
\ref{3frs34b}).

\subsection{Second-order deformation\label{secondord}}

In this section we investigate the consistency of the first-order
deformation, described by equation (\ref{3frs18c}). Along the same line as
before, we can write the second-order deformation like the sum between the
Rarita-Schwinger contribution and the interacting part
\begin{equation}
S_{2}^{\mathrm{A,\psi }}=\tilde{S}_{2}^{\mathrm{A-\psi }}+\tilde{S}_{2}^{%
\mathrm{\psi }}.  \label{3frs64}
\end{equation}%
The piece $\tilde{S}_{2}^{\mathrm{\psi }}$ is subject to the equation
\begin{equation}
\frac{1}{2}\left( S_{1},S_{1}\right) ^{\mathrm{\psi }}+s\tilde{S}_{2}^{%
\mathrm{\psi }}=0,  \label{3frs65}
\end{equation}%
where
\begin{equation}
\left( S_{1},S_{1}\right) ^{\mathrm{\psi }}=\left( S_{1}^{\mathrm{\psi }%
},S_{1}^{\mathrm{\psi }}\right) +\left( S_{1}^{\mathrm{A-\psi }},S_{1}^{%
\mathrm{A-\psi }}\right) ^{\mathrm{\psi }}.  \label{3frs65.1}
\end{equation}%
In formula (\ref{3frs65.1}) we used the notation $\left( S_{1}^{\mathrm{%
A-\psi }},S_{1}^{\mathrm{A-\psi }}\right) ^{\mathrm{\psi }}$ for those
pieces from $\left( S_{1}^{\mathrm{A-\psi }},S_{1}^{\mathrm{A-\psi }}\right)
$ that contain only BRST generators from the Rarita-Schwinger spectrum. The
component $\tilde{S}_{2}^{\mathrm{A-\psi }}$ results as solution to the
equation
\begin{equation}
\frac{1}{2}\left( S_{1},S_{1}\right) ^{\mathrm{A-\psi }}+s\tilde{S}_{2}^{%
\mathrm{A-\psi }}=0,  \label{3frs65.2}
\end{equation}%
where
\begin{equation}
\left( S_{1},S_{1}\right) ^{\mathrm{A-\psi }}=2\left( S_{1}^{\mathrm{\psi }%
},S_{1}^{\prime \mathrm{A-\psi }}\right) +\left( S_{1}^{\prime \mathrm{%
A-\psi }},S_{1}^{\prime \mathrm{A-\psi }}\right) ^{\mathrm{A-\psi }}
\label{3frs65.3}
\end{equation}%
and $\left( S_{1}^{\prime \mathrm{A-\psi }},S_{1}^{\prime \mathrm{A-\psi }%
}\right) ^{\mathrm{A-\psi }}=\left( S_{1}^{\prime \mathrm{A-\psi }%
},S_{1}^{\prime \mathrm{A-\psi }}\right) -\left( S_{1}^{\mathrm{A-\psi }%
},S_{1}^{\mathrm{A-\psi }}\right) ^{\mathrm{\psi }}$. If we denote by $%
\tilde{\Delta}^{\mathrm{\psi }}$ and $\tilde{b}^{\mathrm{\psi }}$ the
nonintegrated densities of the functionals $\left( S_{1},S_{1}\right) ^{%
\mathrm{\psi }}$ and respectively $\tilde{S}_{2}^{\mathrm{\psi }}$, then the
local form of (\ref{3frs65}) becomes
\begin{equation}
\tilde{\Delta}^{\mathrm{\psi }}=-2s\tilde{b}^{\mathrm{\psi }}+\partial _{\mu
}\tilde{n}^{\mu },  \label{3frs67}
\end{equation}%
with
\begin{equation}
\mathrm{gh}\left( \tilde{\Delta}^{\mathrm{\psi }}\right) =1,\qquad \mathrm{gh%
}\left( \tilde{b}^{\mathrm{\psi }}\right) =0,\qquad \mathrm{gh}\left( \tilde{%
n}^{\mu }\right) =1,  \label{3frs68}
\end{equation}%
for some local currents $n^{\mu }$. Direct computation shows that $\tilde{\Delta} ^{%
\mathrm{\psi }}$ decomposes as
\begin{equation}
\tilde{\Delta}^{\mathrm{\psi }}=\tilde{\Delta}_{1}^{\mathrm{\psi }}+\tilde{%
\Delta}_{0}^{\mathrm{\psi }},\qquad \mathrm{agh}\left( \tilde{\Delta}_{1}^{%
\mathrm{\psi }}\right) =1,\qquad \mathrm{agh}\left( \tilde{\Delta}_{0}^{%
\mathrm{\psi }}\right) =0,  \label{we1}
\end{equation}%
where
\begin{eqnarray}
\tilde{\Delta}_{1}^{\mathrm{\psi }} &=&\partial _{\mu }\tilde{\tau}_{1}^{\mu
}+\gamma \left[ -\frac{\mathrm{i}\tilde{k}^{2}}{3}\left( \psi _{\lbrack \mu
}^{\ast }\gamma _{\nu \rho \lambda ]}^{\left. {}\right. }\xi -\frac{1}{2}%
\psi ^{\ast \sigma }\gamma _{\mu \nu \rho \lambda \sigma }\xi \right) \bar{%
\psi}^{\mu }\gamma ^{\nu \rho }\psi ^{\lambda }\right]  \notag \\
&&-\frac{\mathrm{i}\tilde{k}^{2}}{3}\left( \psi _{\lbrack \mu }^{\ast
}\gamma _{\nu \rho \lambda ]}^{\left. {}\right. }\xi -\frac{1}{2}\psi ^{\ast
\sigma }\gamma _{\mu \nu \rho \lambda \sigma }\xi \right) \bar{\xi}\gamma
^{\mu \nu }\partial ^{\lbrack \rho }\psi ^{\lambda ]},  \label{3frs72}
\end{eqnarray}%
\begin{eqnarray}
\tilde{\Delta}_{0}^{\mathrm{\psi }} &=&\partial _{\mu }\tilde{\tau}_{0}^{\mu
}+180\mathrm{i}m^{2}\bar{\xi}\gamma ^{\mu }\psi _{\mu }  \notag \\
&&+\mathrm{i}\tilde{k}^{2}\left( \bar{\psi}_{[\mu }\gamma _{\nu \rho }\psi
_{\lambda ]}+\frac{1}{2}\bar{\psi}^{\alpha }\gamma _{\alpha \beta \mu \nu
\rho \lambda }\psi ^{\beta }\right) \partial ^{\mu }\left( \bar{\xi}\gamma
^{\nu \rho }\psi ^{\lambda }\right)  \label{3frs72a}
\end{eqnarray}

Because $\left( S_{1},S_{1}\right) ^{\mathrm{\psi }}$ contains terms of
maximum antighost number equal to one, we can assume (without loss of
generality) that $\tilde{b}^{\mathrm{\psi }}$ stops at antighost number two%
\begin{eqnarray}
\tilde{b}^{\mathrm{\psi }} &=&\sum\limits_{I=0}^{2}\tilde{b}_{I}^{\mathrm{%
\psi }},\qquad \mathrm{agh}\left( \tilde{b}_{I}^{\mathrm{\psi }}\right)
=I,\qquad I=\overline{0,2},  \label{3frs69a} \\
\tilde{n}^{\mu } &=&\sum\limits_{I=0}^{2}\tilde{n}_{I}^{\mu },\qquad \mathrm{%
agh}\left( \tilde{n}_{I}^{\mu }\right) =I,\qquad I=\overline{0,2}.
\label{3frs69b}
\end{eqnarray}%
By projecting equation (\ref{3frs67}) on the various (decreasing) values of
the antighost number, we then infer the equivalent tower of equations
\begin{eqnarray}
0 &=&-2\gamma \tilde{b}_{2}^{\mathrm{\psi }}+\partial _{\mu }\tilde{n}%
_{2}^{\mu },  \label{3frs70a} \\
\tilde{\Delta}_{I}^{\mathrm{\psi }} &=&-2\left( \delta \tilde{b}_{I+1}^{%
\mathrm{\psi }}+\gamma \tilde{b}_{I}^{\mathrm{\psi }}\right) +\partial _{\mu
}\tilde{n}_{I}^{\mu },\qquad I=\overline{0,1}.  \label{3frs70b}
\end{eqnarray}%
Equation (\ref{3frs70a}) can always be replaced with
\begin{equation}
\gamma \tilde{b}_{2}^{\mathrm{\psi }}=0.  \label{a74xz}
\end{equation}%
Thus, $\tilde{b}_{2}^{\mathrm{\psi }}$ belongs to the Rarita-Schwinger
sector of cohomology of $\gamma $, $H\left( \gamma \right) $. By means of
definitions (\ref{3frs14e})--(\ref{3frs14g}) we get that $H\left( \gamma
\right) $ in the Rarita-Schwinger sector is generated by the objects $\left(
\psi ^{\ast \mu },\xi ^{\ast },\partial _{\lbrack \mu }\psi _{\nu ]}\right) $%
, by their spacetime derivatives up to a finite order, and also by the
undifferentiated ghosts $\xi $ (the spacetime derivatives of $\xi $ are $%
\gamma $-exact according to the second relation in (\ref{3frs14f})). As a
consequence, we can write
\begin{equation*}
\tilde{b}_{2}^{\mathrm{\psi }}=\tilde{\beta}_{2}^{\mathrm{\psi }}\left( %
\left[ \partial _{\lbrack \mu }\psi _{\nu ]}\right] ,\left[ \psi ^{\ast \mu }%
\right] ,\left[ \xi ^{\ast }\right] \right) e^{2}\left( \xi \right) ,
\end{equation*}%
where $e^{2}\left( \xi \right) $ are the elements of pure ghost number two
of a basis in the space of polynomials in the ghosts $\xi $, (\ref{3frs50}).

We observe that $\tilde{\Delta}_{1}^{\mathrm{\psi }}$ from (\ref{3frs72})
can be written as in (\ref{3frs70b}) for $I=1$ if and only if
\begin{equation}
\tilde{\chi}=-\frac{\mathrm{i}\tilde{k}^{2}}{3}\left( \psi _{\lbrack \mu
}^{\ast }\gamma _{\nu \rho \lambda ]}^{\left. {}\right. }\xi -\frac{1}{2}%
\psi ^{\ast \sigma }\gamma _{\mu \nu \rho \lambda \sigma }\xi \right) \bar{%
\xi}\gamma ^{\mu \nu }\partial ^{\lbrack \rho }\psi ^{\lambda ]}  \label{hi}
\end{equation}%
reads as
\begin{equation}
\tilde{\chi}=-2\delta \tilde{b}_{2}^{\mathrm{\psi }}+\gamma \tilde{\rho}%
+\partial _{\mu }\tilde{l}^{\mu },  \label{rephi}
\end{equation}%
where
\begin{equation}
\tilde{\rho}=\frac{\mathrm{i}\tilde{k}^{2}}{3}\left( \psi _{\lbrack \mu
}^{\ast }\gamma _{\nu \rho \lambda ]}^{\left. {}\right. }\xi -\frac{1}{2}%
\psi ^{\ast \sigma }\gamma _{\mu \nu \rho \lambda \sigma }\xi \right) \bar{%
\psi}^{\mu }\gamma ^{\nu \rho }\psi ^{\lambda }-2\tilde{b}_{1}^{\mathrm{\psi
}}.  \label{mn1}
\end{equation}%
Assume that (\ref{rephi}) holds. Then, by taking its left Euler-Lagrange
(EL) derivatives with respect to $\psi _{\mu }^{\ast }$ and using the
commutation between $\gamma $ and each EL derivative $\delta ^{\mathrm{L}%
}/\delta \psi _{\mu }^{\ast }$, we infer the relations
\begin{equation}
\frac{\delta ^{\mathrm{L}}\left( \tilde{\chi}+2\delta \tilde{b}_{2}^{\mathrm{%
\psi }}\right) }{\delta \psi _{\mu }^{\ast }}=\gamma \left( \frac{\delta ^{%
\mathrm{L}}\tilde{\rho}}{\delta \psi _{\mu }^{\ast }}\right) .
\label{pfrs4.21e}
\end{equation}%
As $\tilde{b}_{2}^{\mathrm{\psi }}$ is $\gamma $-invariant, then $\delta
\tilde{b}_{2}^{\mathrm{\psi }}$ will also be $\gamma $-invariant. Recalling
the previous results on the cohomology of $\gamma $ in the Rarita-Schwinger
sector, we find that $\delta \tilde{b}_{2}^{\mathrm{\psi }}=e^{2}\left( \xi
\right) \psi _{\mu }^{\ast }v^{\mu }$, with $v^{\mu }$ fermionic, $\gamma $%
-invariant functions of antighost number zero and $e^{2}\left( \xi \right) $
the elements of pure ghost number two of a basis in the space of polynomials
in the ghosts $\xi $ . By using (\ref{hi}) and the last expression of $%
\delta \tilde{b}_{2}^{\mathrm{\psi }}$, direct computation provides the
equation
\begin{eqnarray}
&&\frac{\delta ^{\mathrm{L}}\left( \tilde{\chi}+2\delta \tilde{b}_{2}^{%
\mathrm{\psi }}\right) }{\delta \psi _{\mu }^{\ast }}=\frac{\mathrm{i}\tilde{%
k}^{2}}{3}\left[ -\frac{2}{3}\left( \gamma _{\nu \rho \lambda }\xi \right)
\left( \bar{\xi}\gamma ^{\lbrack \mu \nu }\partial ^{\rho }\psi ^{\lambda
]}\right) \right.  \notag \\
&&\left. +\frac{1}{2}\left( \gamma ^{\mu \nu \rho \lambda \sigma }\xi
\right) \left( \bar{\xi}\gamma _{\nu \rho }\partial _{\lbrack \lambda }\psi
_{\sigma ]}\right) \right] +2e^{2}\left( \xi \right) v^{\mu }.  \label{x}
\end{eqnarray}%
On the one hand, equation (\ref{pfrs4.21e}) shows that $\delta ^{\mathrm{L}%
}\left( \tilde{\chi}+2\delta \tilde{b}_{2}^{\mathrm{\psi }}\right) /\delta
\psi _{\mu }^{\ast }$ is trivial in $H\left( \gamma \right) $. On the other
hand, relation (\ref{x}) emphasizes that $\delta ^{\mathrm{L}}\left( \tilde{%
\chi}+2\delta \tilde{b}_{2}^{\mathrm{\psi }}\right) /\delta \psi
_{\mu }^{\ast }$ is a nontrivial element from $H\left( \gamma
\right) $ (because each term on the right-hand side of (\ref{x}) is
nontrivial in $H\left( \gamma \right) $). Then, $\delta
^{\mathrm{L}}\left( \tilde{\chi}+2\delta \tilde{b}_{2}^{\mathrm{\psi
}}\right) /\delta \psi _{\mu }^{\ast }$ must be set zero
\begin{equation}
\frac{\delta ^{\mathrm{L}}\left( \tilde{\chi}+2\delta \tilde{b}_{2}^{\mathrm{%
\psi }}\right) }{\delta \psi _{\mu }^{\ast }}=0,  \label{mn2}
\end{equation}%
which yields\footnote{%
In fact, the general solution to equation (\ref{mn3}) takes the form $\tilde{%
\chi}+2\delta \tilde{b}_{2}^{\mathrm{\psi }}=u+\partial _{\mu }\tilde{l}%
^{\mu }$, where $u$ is a function of antighost number one depending on all
the BRST generators from the Rarita-Schwinger sector but the antifields $%
\psi _{\mu }^{\ast }$. As the antifields $\psi _{\mu }^{\ast }$ are the only
Rarita-Schwinger antifields of antighost number one, the condition $\mathrm{%
agh}\left( u\right) =1$ automatically produces $u=0$.}
\begin{equation}
\tilde{\chi}+2\delta \tilde{b}_{2}^{\mathrm{\psi }}=\partial _{\mu }\tilde{l}%
^{\mu }.  \label{mn3}
\end{equation}%
By acting with $\delta $ on (\ref{mn3}) we deduce
\begin{equation}
\delta \tilde{\chi}=\partial _{\mu }\tilde{j}^{\mu }.  \label{mn4}
\end{equation}%
From (\ref{hi}), by direct computation we find
\begin{equation}
\delta \tilde{\chi}=\frac{\tilde{k}^{2}}{3}\left( \partial ^{\alpha }\bar{%
\psi}^{\beta }\right) \gamma _{\alpha \beta \mu }\left[ \frac{2}{3}\left(
\gamma _{\nu \rho \lambda }\xi \right) \bar{\xi}\gamma ^{\lbrack \mu \nu
}\partial ^{\rho }\psi ^{\lambda ]}-\frac{1}{2}\left( \gamma ^{\mu \nu \rho
\lambda \sigma }\xi \right) \bar{\xi}\gamma _{\nu \rho }\partial _{\lbrack
\lambda }\psi _{\sigma ]}\right] .  \label{mn5}
\end{equation}%
Comparing (\ref{mn4}) with (\ref{mn5}) and recalling the Noether identities
corresponding to the Rarita-Schwinger action, we obtain that the right-hand
of (\ref{mn5}) reduces to a total derivative iff
\begin{equation}
\frac{2}{3}\left( \gamma _{\nu \rho \lambda }\xi \right) \bar{\xi}\gamma
^{\lbrack \mu \nu }\partial ^{\rho }\psi ^{\lambda ]}-\frac{1}{2}\left(
\gamma ^{\mu \nu \rho \lambda \sigma }\xi \right) \bar{\xi}\gamma _{\nu \rho
}\partial _{\lbrack \lambda }\psi _{\sigma ]}=\partial ^{\mu }\tilde{p}.
\label{mn6}
\end{equation}%
Simple computation exhibits that the left-hand side of (\ref{mn6}) cannot be
written like a total derivative, so neither relation (\ref{mn4}) nor
equation (\ref{rephi}) hold. As a consequence, $\tilde{\chi}$ must vanish
and hence we must set
\begin{equation}
\tilde{k}=0.  \label{pfrs4.22}
\end{equation}

Inserting (\ref{pfrs4.22}) in (\ref{3frs72})--(\ref{3frs72a}), we obtain
that
\begin{eqnarray}
\tilde{\Delta}_{1}^{\mathrm{\psi }} &=&\partial _{\mu }\tilde{\tau}_{1}^{\mu
},  \label{pfrs4.33a} \\
\tilde{\Delta}_{0}^{\mathrm{\psi }} &=&\partial _{\mu }\tilde{\tau}_{0}^{\mu
}+180\mathrm{i}m^{2}\bar{\xi}\gamma ^{\mu }\psi _{\mu }.  \label{pfrs4.33b}
\end{eqnarray}
From (\ref{pfrs4.33a}) it results that we can safely take $\tilde{b}_{2}^{%
\mathrm{\psi }}=0$ and $\tilde{b}_{1}^{\mathrm{\psi }}=0$, which replaced in
(\ref{pfrs4.33b}) lead to the necessary condition that $\tilde{\Delta}_{0}^{%
\mathrm{\psi }}$ must be a trivial element from the local cohomology of $%
\gamma $, i.e. $\tilde{\Delta}_{0}^{\mathrm{\psi }}=-2\gamma \tilde{b}_{0}^{%
\mathrm{\psi }}+\partial _{\mu }\tilde{n}_{0}^{\mu }$. In order to solve
this equation with respect to $\tilde{b}_{0}^{\mathrm{\psi }}$, we will
project it on the number of derivatives. Since $\gamma \tilde{b}_{0}^{%
\mathrm{\psi }}$ contains at least one spacetime derivative, the above
equation projected on the number of derivatives equal to zero reduces to $%
\tilde{\Delta}_{0}^{\mathrm{\psi }}=180\mathrm{i}m^{2}\bar{\xi}\gamma ^{\mu
}\psi _{\mu }=0$, which further implies
\begin{equation}
m=0,  \label{fg3}
\end{equation}
so
\begin{equation}
S_{1}^{\mathrm{\psi }}=0.  \label{fg4}
\end{equation}
Replacing (\ref{fg3}) and (\ref{fg4}) in (\ref{def1}), we obtain that the
general form of the first-order deformation for the free model under study
that is consistent to the second order in the coupling constant reads as
\begin{equation}
S_{1}^{\mathrm{A,\psi }}=S_{1}^{\mathrm{A}}\equiv q\int d^{11}x\,\varepsilon
^{\mu _{1}\ldots \mu _{11}}A_{\mu _{1}\mu _{2}\mu _{3}}F_{\mu _{4}\ldots \mu
_{7}}F_{\mu _{8}\ldots \mu _{11}}.  \label{fg5}
\end{equation}
Inserting (\ref{fg5}) into (\ref{3frs18b})--(\ref{3frs18d}), etc., we find
that all the higher-order deformations can be taken to vanish
\begin{equation}
S_{k}=0,\qquad k>1,  \label{fg6}
\end{equation}
so the full deformed solution to the master that is consistent to all orders
in the coupling constant takes the simple form
\begin{equation}
\bar{S}=S+\lambda S_{1}^{\mathrm{A}},  \label{fg7}
\end{equation}
where $S$ is the solution to the master equation for the starting free
model, (\ref{3frs15}). Relation (\ref{fg7}) emphasizes that under the
hypotheses mentioned at the beginning of this section, there are neither
cross-couplings that can be added between an Abelian three-form gauge field
and a massless gravitino nor self-interactions for the gravitino in $D=11$.

\section{Conclusion\label{conc}}

To conclude with, in this paper we have investigated the consistent
interactions in eleven spacetime dimensions that can be added to a
free theory describing a massless gravitino and an Abelian $3$-form
gauge field. Our treatment is based on the Lagrangian BRST
deformation procedure, which relies on the construction of
consistent deformations of the solution to the master equation with
the help of standard cohomological techniques. We worked under the
hypotheses of smoothness in the coupling constant, locality, Lorentz
covariance, Poincar\'{e} invariance, and the preservation of the
number of derivatives on each field. Our main result is that there
are neither cross-couplings that can be added between an Abelian
three-form gauge field and a massless gravitino nor
self-interactions for the gravitino in $D=11$. The only possible
term that can be added to the deformed solution to the master
equation is nothing but a generalized Chern-Simons term for the
three-form gauge field, (\ref{fg5}), which brings contributions to
the deformed Lagrangian, but does not modify the original, Abelian
gauge transformations (\ref{3frs8}).

\section*{Acknowledgments}

The authors wish to thank Constantin Bizdadea and Odile Saliu for useful
discussions and comments. This work is partially supported by the European
Commission FP6 program MRTN-CT-2004-005104 and by the grant AT24/2005 with
the Romanian National Council for Academic Scientific Research
(C.N.C.S.I.S.) and the Romanian Ministry of Education and Research (M.E.C.).

\end{document}